\documentstyle[12pt,amstex,amssymb,righttag]{article}

\begin{document}
%\renewcommand{\thefootnote}{\fnsymbol{footnote}}
%\special{papersize=210mm,297mm}

\thispagestyle{empty}

\vspace{3cm}

\begin{center}
{\Large \bf Quantum Gravity in $D=5$ Dimensions}\\[0.5cm]

\vspace{3mm}
by

\vspace{3mm}
{\sl Carlos Pinheiro}$^{\ast}$\footnote{fcpnunes@@cce.ufes.br/maria@@gbl.com.br} \\[0.5cm]
and \\[0.5cm]
{\sl F.C. Khanna}$^{\ast \ast}$\footnote{khanna@@phys.ualberta.ca}

\vspace{3mm}
$^{\ast}$Universidade Federal do Esp\'{\i}rito Santo, UFES.\\
Centro de Ci\^encias Exatas\\
Av. Fernando Ferrari s/n$^{\underline{0}}$\\
Campus da Goiabeiras 29060-900 Vit\'oria ES -- Brazil.\\

$^{\ast \ast}$Theoretical Physics Institute, Dept. of Physics\\
University of Alberta,\\
Edmonton, AB T6G2J1, Canada\\
and\\
TRIUMF, 4004, Wesbrook Mall,\\
V6T2A3, Vancouver, BC, Canada.
\end{center}

\vspace{3mm}
\begin{center}
Abstract
\end{center}

We propose a topological Chern-Simons term in $D=5$ dimensions
coupled to
Einstein Hilbert theory. Hartree approximation for topological
Lagrangian and the Chern-Simons term in $D=3$ is considered. An
effective model of Quantum Gravity in $D=5$ dimensions is presented
here. The analysis of residues is considered and the unitarity is
guaranteed at tree level. The propagator is ghost  and tachyon free.

\newpage
\subsection*{Introduction}

\paragraph*{}
It's known that the theory of Quantum Gravity has numerous problems.
In particular the perturbative approach contains the insoluble 
conflict between unitarity and renormalizability in $D=4$ dimensions. 
The Einstein Hilbert theory, for example, is unitarity but is not
renormalizable. Anyway the theory can be at least seen as an
effective theory. 
Many models for perturbative quantum gravity have been constructed in
$D=4$ including theories with higher derivatives\cite{quatro}.
Another interesting development occured with the discovery of
topological Chern-Simons term.
That term despite having some important geometric properties does't
have any physics associated with it.
The Chern-Simons term in $D=3$ dimensions together with
Einstein-Hilbert Lagrangian provides a rich array of physics.
All problems of Quantum Gravity in $D=4$ disappear in $D=3$. The so
called topological theory for gravitation or still the
Einstein-Hilbert-Chern-Simons theory is unitarity and finite\cite{dois,quatro}.

It can be shown that the dynamics is given by a massive pole and the
massless pole does't have propagation in $D=3$ \cite{quatro}.

Presently we know that in general it is possible to have a
topological term like Chern-Simons in odd dimensions\cite{seis}, $D=3,5,7\cdots$.

The question, however is what we might do to make a perturbative
attack on Einstein-Chern-Simons in $D=3$, but may we do the same approach for
$D=5$ dimensions?

Is it possible to carry out a perturbative approach in $5$-dimensions
as was the case for $D=3$?

To answer this question a convenient ``Chern-Simons'' in $D=5$ is
needed and then to establish that a perturbative approach on a
background is possible.

It is possible \cite{um} to find Chern-Simons term in $D=5$
but there is no indication of a perturbative approach to the problem.
There is nothing to describe the free theory for gravitation in $D=5$
as in the case of $D=3$. It is speculated that ``Chern-Simons'' in
$D=5$ is self interacting and so would be impossible to write in
analogy to the Chern-Simons term in $D=3$, where is possible to have a
part that describes the free theory and another part that describes
the gravitational interaction.

For calculation of the propagator in perturbation theory the choice of
background is shown to be important.

Some possible topological terms in $D=5$ go to zero as the perturbation on a
background is introduced. For example in a flat space time
background, $\eta_{\mu \nu}$, the bilinear term in the
topological Lagrangian, is not possible.

In $D=3$ we can construct a topological free theory for gravitation
on flat space time if we suppose that the field variable $h_{\mu
\nu}(x)$ tranforms like a tensor.

We wish to do a similar treatment for an
Einstein-Hilbert-``Chern-Simons'' theory in $D=5$.

We propose a topological Lagrangian in $D=5$ similar to the one given
in \cite{quatro}. Since
the analysis in $D=3$ are made with only the first part of the
topological Lagrangian with the primary interest being in a free
theory; only a part of the topological Lagrangian in $D=5$ is written
here and the coupled Einstein-Hilbert Lagrangian is considered.

The Chern-Simons term in $D=3$ has a global invariance by
diffeomorphism, but the local invariance is guaranteed because
$h_{\mu\nu}$ transforms like a tensor.

The global covariance by diffeomorphism for ``Chern-Simons'' in $D=5$
is not known. The analogy from first part of Chern-Simons in $D=3$ is
used, but it is assumed that the second term exists and that the
local covariance is assumed, since as before $h_{\mu\nu}$ is a tensor.

Assume a topological term in $D=5$ and a Hartree approximation for 
our topological term and Chern-Simons in $D=3$ is considered. 

Finally, the calculation of the propagator and an analysis of unitarity
in tree level is carried out. The theory is seen as an effective
theory. Thus there is no problem with renormalizability.

The Lagrangian for Einstein-Chern-Simons theory in $D=5$ dimensions
is given as  
\begin{equation}
{\cal L}={\cal L}_{{}_{E\cdot H}}+{\cal L}_{g\cdot f}+{\cal L}_{c\cdot s}
\end{equation}
where ${\cal L}_{{}_{E\cdot H}},\ {\cal L}_{g\cdot f}$ and ${\cal
L}_{c\cdot s}$ 
are respectively the Einstein Hilbert Lagrangian, gauge fixing
Lagrangian and the topological Chern-Simons term in $D=5$ dimensions.
These are 
\begin{eqnarray}
{\cal L}_{{}_{E\cdot H}} &=& -\frac{1}{2k^2}\ \sqrt{-g}\ R\ , \\
{\cal L}_{g\cdot f} &=& \frac{1}{2\alpha}\ F_{\mu}F^{\mu}\ , \\
{\cal L}_{c\cdot h} &=& \frac{1}{\mu}\ \varepsilon^{\mu\nu\alpha\beta\gamma}
\varepsilon^{\lambda \theta \xi \Delta}\tau \partial_{\mu}
\Gamma^{w}_{\lambda \nu}\partial_{\mbox{{\footnotesize  k}}}
\Gamma^{\mbox{{\footnotesize  k}}}_{\alpha \theta}\partial_{\psi}
\Gamma^{\psi}_{\beta \xi}\partial_{w}\Gamma^{\tau}_{\gamma \Delta}\ .
\end{eqnarray}

In eq. (2) $k$ is the gravitational constant. $R$ is the ususal
scalar curvature and $\sqrt{-g}$ is the determinant of the metric.

In  equation (3) $F_{\mu}$ represents the De Donder gauge fixing
term given by
\begin{equation}
F_{\mu}\left[h_{\rho \sigma}\right]=\partial_{\lambda}
\left(h^{\lambda}_{\mu}-\frac{1}{2}\ \delta^{\lambda}_{\mu}
h^{\nu}_{\nu}\right)
\end{equation}
and $\underline{\alpha}$ is the Feymann parameter. In eq. (4)
$\varepsilon^{\mu\nu\alpha\beta\gamma}$ and
$\Gamma^{\alpha}_{\mu\nu}$ are Levi-Civita and Christoffel symbols respectively.

The gauge fixing invariance is expressed as
\begin{equation}
\delta h_{\mu\nu}(x)=\partial_{\mu}\xi_{\nu}(x)+
\partial_{\nu}\xi_{\mu}(x)
\end{equation}

The pertubation theory on flat space time is considered such that
\begin{equation}
g_{\mu\nu}(x)=\eta_{\mu\nu}+kh_{\mu\nu}(x)
\end{equation}
where $h_{\mu\nu}(x)$ will be the gravitation field variable.

Then eq. (1) has a bilinear form like $h\theta h$, and $\theta$ is an
operator associated with the spin projection operators in rank-2
tensor space.

The Chern-Simons Lagrangian in $D=5$ dimension is not of the bilinear
form in $h_{\mu\nu}$, but a square bilinear like $hh\theta^{\ast}hh$.
The Lagrangian can be written as
\begin{equation}
{\cal L}=h^{\mu\nu}\theta_{\mu\nu ,k\lambda}h^{k\lambda}+
h^{w\chi}h^{k\delta}\theta_{w\chi k\delta ,\psi\sigma \zeta \tau}h^{\psi \sigma}
h^{\zeta \tau}\ .
\end{equation}

With this form it is difficult to find the propagator.
However use of the Hartree approximation \cite{cinco} between Chern-Simons in
$D=5$ dimensions and Chern-Simons in $D=3$ dimensions is possible.

Chern-Simons in $D=3$ is given by
\begin{equation}
{\cal L}=\varepsilon^{\mu\nu\alpha}\left(\Gamma^{\lambda}_{\mu\beta}
\partial_{\nu}\Gamma^{\beta}_{\alpha \lambda}+\frac{2}{3}\ 
\Gamma^{\rho}_{\mu \sigma}\Gamma^{\sigma }_{\nu \varphi}\Gamma^{\varphi}_{\alpha \rho}\right)
\end{equation}

If Hartree approximation is assumed it can be shown that 
\[
{\cal L}_{c.h}(D=5)\simeq \lambda^2{\cal L}_{c.h}(D=3)
\]
where the left side means the Chern-Simons in $D=5$, the right side
is the Chern-Simons in $D=3$ dimensions.

The parameter $\lambda^2$ is a real parameter and it will describe
the physics in the model since some conditions are necessary to
achieve unitarity of the theory in the tree level. Essentially what we
are doing is to consider the square bilinear $hh\theta^{\ast}hh$
as an approximation described by a real parameter times $h\theta h$,
where $h\theta h$ is the linearized Chern-Simons  in $D=3$.

Then in the Hartree approximation 
\begin{equation}
{\cal L}=\frac{1}{2}\ h^{\mu\nu}
\left(\theta_{\mu\nu ,k\lambda} \oplus  \theta_{\mu\nu
,k\lambda}^{\ast}\right)h^{k\lambda}
\end{equation}
where $\theta_{\mu\nu ,k\lambda}$ is the contribution from the
Einstein-Hilbert Lagrangian including the gauge fixing and
$\theta^{\ast}_{\mu\nu ,k\lambda}$ is the operator generated by the
topological Chern-Simons term in $D=3$. 

The operators $\theta$ and $\theta^{\ast}$ are given respectively by
\begin{equation}
\theta_{\mu\nu ,k\lambda}=\frac{\Box}{2}\ \overset{(2)}{P}+
\frac{\Box}{2\alpha}\ \overset{(1)}{P}_{\!\!m}-\Box\left(
\frac{4\alpha -3}{4\alpha}\right)\overset{(0)}{P}_{\!\!s}+
\frac{\Box}{4\alpha}\ \overset{(0)}{P}_{\!\!w}-
\frac{\Box\sqrt{3}}{4\alpha}\ \overset{(0)}{P}_{\!\!sw}-
\frac{\Box\sqrt{3}}{\alpha}\ \overset{(0)}{P}_{\!\!ws}
\end{equation}
and
\begin{equation}
\theta^{\ast}_{\mu \nu k\lambda}=\frac{4k^2}{\mu}\
\lambda^2(S_1+S_2)\ .
\end{equation}

Here $\overset{(2)}{P},\ \overset{(1)}{P}_{\!\!m}, \
\overset{(0)}{P}_{\!\!s},\ \overset{(0)}{P}_{\!\!w},\ \overset{(0)}{P}_{\!\!sw}$
and $\overset{(0)}{P}_{sw}$ are spin projection operators.

The two new operators are $S_1$ and $S_2$  and are given by.
\begin{eqnarray}
(S_1)_{\mu\nu k\lambda} &=& \frac{1}{4}(-\Box )\left[
\varepsilon_{\mu\alpha\lambda}\partial_kw^{\alpha}_{\nu}+
\varepsilon_{\mu\nu k}\partial_{\lambda}w^{\alpha}_{\nu}+
\varepsilon_{\nu\alpha\lambda}\partial_kw^{\alpha}_{\mu}+
\varepsilon_{\nu\alpha
k}\partial_{\lambda}w^{\alpha}_{\mu}\right] \\
(S_2)_{\mu\nu k\lambda} &=& \frac{1}{4}\Box \big[
\varepsilon_{\mu\alpha\lambda}\eta_{k\nu}+
\varepsilon_{\mu\alpha k}\eta_{\lambda \nu}+
\varepsilon_{\nu\alpha\lambda}\eta_{k\mu}+
\varepsilon_{\nu\alpha k}\eta_{\lambda \mu}\big]\partial^{\alpha}\ .
\end{eqnarray}
We are looking for the propagator of Einstein Hilbert-Chern-Simons in
$D=5$ dimensions, then we assume a linear combination of the same spin
projection operators
\begin{equation}
(\theta^{-1})_{\mu\nu k\lambda}=X\overset{(2)}{P}+
Y\overset{(1)}{P}_{\!\!m}+Z\overset{(0)}{P}_{\!\!s}+
W\overset{(0)}{P}_{\!\!w}+T\overset{(0)}{P}_{\!\!sw}+R\overset{(2)}{P}_{\!\!ws}+
MS_1+NS_2  \ .
\end{equation}

Now we can calculate the propagator for the field $h_{\mu \nu}(x)$ in
$D=5$ by extending the algebra of Barnes and Rivers \cite{quatro} and the
inverse operator given by eq. (15).

When we take the complete operator from eq. (10) and the inverse operator in
eq. (15), the multiplication between them give us the identity in rank-2
tensor space, as 
\begin{equation}
\theta_{\mu\nu}^{\hspace*{3mm}\rho\sigma}\left(\theta_{\rho\sigma k\lambda}\right)^{-1}=
\left(\overset{(2)}{P}+\overset{(1)}{P}_{\!\!m}+\overset{(0)}{P}_{\!\!s}+
\overset{(0)}{P}_{\!\!w}\right)_{\mu \nu ,k\lambda}\ .
\end{equation}

A system of equation are found from eq. (16) and these are
\begin{eqnarray}
&{}&\frac{\Box}{2}\ X - \frac{4k^2\lambda^2}{\mu}\ \Box^3N=1\ , \nonumber \\
&&\nonumber \\
&{}&\frac{\Box}{2\alpha}\ Y = 1\ , \nonumber \\
&&\nonumber \\
&{}&\Big[\frac{\Box}{6}+\frac{\Box}{3}\left(\frac{4\alpha -3}{4\alpha}\right)\Big]
X + 
\left[-\frac{\Box}{6}-\Box \left(\frac{4\alpha -3}{3\alpha}\right)\right]Z-
\frac{\Box\sqrt{3}}{4\alpha}\ R+
\frac{2k^2\lambda^2\Box^3}{\mu}\ N=1\nonumber \\
&&\nonumber \\
&{}&\left(-\frac{\Box}{6}-\frac{\Box\sqrt{3}}{3\alpha}\right)T+
\frac{\Box}{4\alpha}\ W=1\ , \nonumber \\
&&\nonumber \\
&{}&\Box \left(\frac{4\alpha -3}{3\alpha}\right)T+
\frac{\Box\sqrt{3}}{4\alpha}\ W=0\ , \nonumber\\
&&\nonumber \\
&{}&\frac{\Box}{4\alpha}\ R+\frac{\Box\sqrt{3}}{12\alpha}\ X-
\frac{\Box\sqrt{3}}{3\alpha}\ Z =0\\
&{}&\left(\frac{\Box}{2}-\frac{\Box}{2\alpha}\right)N+
\frac{\Box}{2\alpha}\ M+\frac{4k^2\lambda^2}{\mu}\ X=0\ \mbox{and}\nonumber \\
&&\nonumber \\
&{}&\frac{\Box}{2}\ N +
\frac{4k^2\lambda^2}{\mu}\ X=0\ , \nonumber
\end{eqnarray}
The coeficients $(X,Y,Z,W,T,R,M,N)$ in the space of coordinates are
written as
\begin{eqnarray}
X &=&\frac{-2}{-\Box -\frac{64\Box^2k^4\lambda^4}{\mu^2}}\ , \nonumber \\
Y &=& \frac{2\alpha}{\Box}\ , \nonumber \\
Z &=& \frac{-64k^4\lambda^4}{64\Box k^4\lambda^4+\mu^2}\ , \nonumber \\
W &=& \frac{8(8-\sqrt{3})(-3+4\alpha )}{61\Box}\ , \\
T &=& \frac{6(3-8\sqrt{3})}{61\Box}\ , \nonumber \\
R &=&
\frac{-2(128\Box k^4\lambda^4+\mu^2)}{\sqrt{3}(64\Box^2k^4\lambda^4+\Box
\mu^2)}\ , \nonumber \\
M &=&
\frac{-16k^2\lambda^2\mu}{\Box^2(64\Box k^4\lambda^4+\mu^2)}\ \  \mbox{and}\ \nonumber \\
N &=& \frac{-16k^2\lambda^2\mu}{\Box^2(64\Box k^4\lambda^4+\mu^2)}\ .\nonumber
\end{eqnarray}
The propagator of Einstein-Hilbert-``Chern-Simons'' theory in $D=5$
can be written as 
\begin{equation}
\langle h_{k \nu}(x),h_{\mu\lambda}(y)\rangle
=i\theta^{-1}_{\mu\nu ,k\lambda} \delta^5(x-y)
\end{equation}
We can define the transition amplitude as 
\begin{equation}
{\cal A}=\tau^{\mu\nu\ast}(x)\langle h_{\mu\nu}(x),\ 
h_{k\lambda}(y)\rangle \tau^{k\lambda}(y)
\end{equation}

The coupling between propagator and external currents like energy
momentum tensor is compatible with the gauge symmetry eq. (6).

Several coefficients in (18) vanish due to the transversality
relation \cite{quatro}. Only three coefficients survive and are referred as 
$X$, $Z$ and $N$.

These coefficients in momentum space are 
\begin{eqnarray}
X &=& \frac{-2}{{\mbox{\bf k}}^2\left(1-\frac{64{\mbox{\bf
k}}^2\lambda^4k^4}{\mu^2}\right)}\ , \nonumber \\
Z &=& \frac{1}{{\mbox{\bf k}}^2\left(1-\frac{\mu^2}{64{\mbox{\bf
k}}^2k^4\lambda^4}\right)}\quad \mbox{and}\nonumber \\
N &=& \frac{\mu}{({\mbox{\bf k}}^2)^3\cdot 4k^2\lambda^2\left(1-
\frac{\mu^2}{{\mbox{\bf k}}^2k^4\lambda^4}\right)}\ .
\end{eqnarray}
In the spin two sector we can see two poles given by
\begin{eqnarray}
{\mbox{\bf k}}^2 &=& 0\quad \mbox{and}\nonumber \\
{\mbox{\bf k}}^2 &=& \left(\frac{\mu}{8k^2\lambda^2}\right)^2\ .
\end{eqnarray}
In the zero spin sector ($Z$ coefficient) and in the topological sector
($N$-coefficient we find the same poles.

Observe that when we put $\mu \rightarrow \infty$ we have the $Z$ and
$N$ coefficients vanishing and the $X$ coefficient is written as 
\begin{equation}
X=-\frac{2}{{\mbox{\bf k}}^2}\ .
\end{equation}
This means that the dominant term of the propagator in $D=5$ when the
contribution from Chern-Simons term is null is compatible with the
result given by \cite{tres,quatro}.

We have pure Einstein theory in $D=5$ dimensions with a propagator 
$\langle h, h\rangle \sim \displaystyle{\frac{1}{{\mbox{\bf k}}^2}}$ similar to
the Einstein case in $D=4$. 

To verify the unitarity of the theory at tree level eq. (19) is
considered in momentum space. The imaginary part of the residues of
the amplitude at each pole lead to the necessary 
unitarity condition.

In momentum space eq. (19) is given by
\begin{equation}
{\cal A} =\tau^{\mu\nu\ast}(\vec{\mbox{\bf k}})\langle h_{\mu\nu}(-
\vec{\mbox{\bf k}}),\ 
h_{k\lambda}(\vec{\mbox{\bf k})}\rangle \tau^{k\lambda}(\vec{\mbox{\bf k}})
\end{equation}

The theory will be free of ghost's if 
\begin{equation}
\hspace*{-1cm}I_m Res{\cal A}\Big|_{\mbox{\bf k}^2=0}\ > \ 0
\end{equation}
and
\begin{equation}
I_m Res{\cal A}\Big|_{\mbox{\bf
k}^2=\left(\frac{\mu}{8k^2\lambda^2}\right)^2}\ > \ 0\ .
\end{equation}
The equations are verified if $|\tau_{k\lambda}|^2<0$, and $\mu$ or
$\lambda^2<0$; or if $\frac{\lambda^2}{\mu}<0$.

The equation (25) will be true if $|\tau_{k\lambda}|^2<0$ and $\mu
>0$; or $\lambda^2>0$; or $\frac{\lambda^2}{\mu}>0$.

On taking $\lambda^2={[-\mu ,0)U(0,+\mu ]}$, we have two possibilities
for propagation of gravitons in the Einstein-``Chern-Simons'' theory
in $D=5$. Both poles are dynamical.

The situation here is different from the pure Einstein theory in
$D=4$ and Einstein-Chern-Simons in $D=3$ dimensions. In pure Einstein
theory, $D=4$ there is only one pole or one massless graviton. The pole
$\mbox{\bf k}^2=0$ has propagation in tree level and the Einstein-Hilbert
Lagrangian is free of ghost's. For the Einstein-Chern-Simons theory
in $D=3$ we have two poles \cite{tres, quatro} given by $\mbox{\bf k}^2=0$ and 
$\mbox{\bf k}^2\left(\frac{\mu}{8k^2}\right)^2$, 
but the dynamics is given by the massive pole. There is no
propagation associated with the massless pole.

By taking $\lambda^2=1$, there is partial information from
Einstein-Chern-Simons in $D=3$, but it should be emphasized that the
propagators in $D=3$ and $D=5$ are different.

Finally, if we consider $\lambda^2=\pm \mu$, in according with 
the range given above, the pole will be located at 
$\mbox{\bf k}^2=\left(\frac{1}{8k^2}\right)^2$ and the dynamic propagation is
guaranteed by unitarity in the tree level. The final result is that the
propagation of Einstein-``Chern-Simons'' theory in $D=5$ dimensions
is completely determined by the massless graviton, because the
gravitational interaction is a large scalar force. The massive pole
has a short range for propagating information.

\subsection*{Conclusions:}

\paragraph*{}
We have constructed an effective model for Einstein-Chern-Simons
theory in $D=5$ dimensions.

This model has two dynamical poles and the unitarity is analyzed in the
tree level. The Lagrangian is free of ghost's and tachyons.

As an objective to treat the problem in perturbative approach, Hartree
approximation was used and a convenient topological term like
Chern-Simons in $D=3$ dimensions was constructed unlike the case of
pure Einstein in $D=4$ and the Einstein-Chern-Simons in $D=3$, here
the propagation is associated with both poles. There is no problem with
the renormalizability since the theory is an effective model for gravity.

\subsection*{Acknowledgements:}

\paragraph*{}
I would like to thank the Department of Physics, University of
Alberta for their hospitality. This work was supported by CNPq
(Governamental Brazilian Agencie for Research.

I would like to thank also Dr. Don N. Page for his kindness and attention
with  me at Univertsity of Alberta.

\end{document}